\documentclass[aps,amsmath,twocolumn,amssymb,floatfix,showpacs,superscriptaddress,nofootinbib,longbibliography]{revtex4-1}
\usepackage{mathtools}
\usepackage{setspace}
\usepackage{braket}
\usepackage[dvipsnames]{xcolor}
\usepackage{float}
\usepackage{subfigure}
\usepackage{graphics}
\usepackage{graphicx}
\usepackage{bm}
\usepackage{amsmath,amssymb,bm}
\usepackage{amsfonts}
 
\usepackage[version=4,arrows=pgf-filled,textfontname=sffamily]{mhchem}
\usepackage{MnSymbol}
\usepackage{braket}
\usepackage{float}
\usepackage{tikz}
\usepackage{etoolbox}
\usepackage{blkarray}
\usepackage{pgffor}
\usepackage{float}
\usepackage{silence}
\usepackage{array}
\usepackage{diagbox}
\usepackage{multirow}
\usepackage{makecell}
\usepackage[colorlinks=true,linktoc=page,linkcolor=BrickRed,citecolor=magenta,urlcolor=purple]{hyperref}

\mathchardef\mhyphen="2D 

\newcommand{\etal}{{\it et al.}}

\newcommand\bea{\begin{eqnarray}}
\newcommand\eea{\end{eqnarray}}
\newcommand\beq{\begin{equation}}  
\newcommand\eeq{\end{equation}}

\usepackage[normalem]{ulem}
\definecolor{lime}{HTML}{A6CE39}
\usepackage{sidecap,tikz}
\DeclareRobustCommand{\orcidicon}{\hspace{-1.0mm}
	\begin{tikzpicture}
		\draw[lime, fill=lime] (0.0,0.0) 
		circle [radius=0.15] 
		node[white] {{\fontfamily{qag}\selectfont \tiny \,ID}};
		\draw[white, fill=white] (-0.0525,0.095) 
		circle [radius=0.007];
	\end{tikzpicture}
	\hspace{-3.0mm}
}

\foreach \x in {A, ..., Z}{\expandafter\xdef\csname orcid\x\endcsname{\noexpand\href{https://orcid.org/\csname orcidauthor\x\endcsname}
		{\noexpand\orcidicon}}
}

\AtBeginDocument{%
	\newwrite\bibnotes
	\def\bibnotesext{Notes.bib}
	\immediate\openout\bibnotes=\jobname\bibnotesext
	\immediate\write\bibnotes{@CONTROL{REVTEX41Control}}
	\immediate\write\bibnotes{@CONTROL{%
			apsrev41Control,author="08",editor="1",pages="1",title="1",year="1"}}
	\if@filesw
	\immediate\write\@auxout{\string\citation{apsrev41Control}}%
	\fi
}%

\usepackage{xcolor}
\definecolor{darkred}{rgb}{2,0,0}
\definecolor{darkgreen}{rgb}{0,0.5,0}
\newcommand{\filledtriangle}[1]{%
  \tikz[baseline={(0,-0.2ex)}]%
    \draw[fill=#1, draw=black, line width=0.2pt]
      (0,0) -- (0.18,0) -- (0.075,0.18) -- cycle;%
}
\DeclareRobustCommand{\filledtriangle}[1]{%
  \tikz[baseline={(0,-0.2ex)}]%
    \draw[fill=#1, draw=black, line width=0.3pt]
      (0,0) -- (0.18,0) -- (0.075,0.18) -- cycle;%
}
\newcommand{\filledcircle}[1]{%
  \tikz[baseline={(0,-0.62ex)}]%
    \draw[fill=#1, draw=black, line width=0.05pt]
      (0,0) circle (0.088cm);%
}
\newcommand{\myfilledsquare}[1]{%
  \text{%
    \tikz[baseline={(0,-0.17ex)}]%
      \draw[fill=#1, draw=black] (0,0) rectangle (0.17,0.17);%
  }%
}
\usetikzlibrary{shapes.geometric}
\newcommand{\mystaroutline}{%
  \text{%
    \tikz[baseline={(0,-0.67ex)}]%
      \node[
        star, star points=5, star point ratio=2.9,
        fill=yellow, draw=black, line width=0.15pt,
        minimum size=0.22cm, inner sep=0pt
      ] at (0,0) {};%
  }%
}

\begin{document}

\newcommand{\Z}{\mathbb{Z}}

\title{Emergent topological phases and coexistence of gapless and spectral-localized Floquet quantum spin Hall states via electron-phonon interaction}

\author{Srijata Lahiri$^\S$\orcidA{}}
\email[]{srijata.lahiri@iitg.ac.in} 

\author{Kuntal Bhattacharyya$^\S$\orcidB{}}
\email[]{kuntalphy@iitg.ac.in}

\author{Saurabh Basu}
\email[]{saurabh@iitg.ac.in}
\affiliation{Department of Physics, Indian Institute of Technology Guwahati, Guwahati-781039, Assam, India}
\def\thefootnote{$\S$}\footnotetext{These authors contributed equally to this work}
\begin{abstract}
In this work, a thorough exploration has been carried out to unravel the role of electron-phonon interaction (EPI) in a Bernevig-Hughes-Zhang (BHZ) quantum spin Hall (QSH) insulator subjected to a time-periodic step drive.
It is observed that upon inclusion of the EPI, the system demonstrates emergent Floquet QSH (FQSH) phases and several topological phase transitions therein, mediated solely by the interaction strength.
Quite intriguingly, the emergence of topological zero ($\pi$) modes in the bulk that remains otherwise gapless in the vicinity of the $\pi$ (zero) energy sector is observed, thus serving as a prime candidate of robust topology in gapless systems.
With other invariants being found to be deficient in characterizing such coexistent phases, a spectral localizer ($\mathcal{SL}$) is employed, which distinctly ascertains the nature of the (zero or $\pi$) edge modes.
Following the $\mathcal{SL}$ prescription, a real-space Chern marker computed by us further provides support to such \textit{gapless} Floquet topological scenario. Our results can be realized in advanced optical setups that may underscore the importance of EPI-induced Floquet features.  
\end{abstract}
\maketitle

\textit{Introduction.} Ever since the field of topological insulators (TIs)~\cite{Kane2005,Bernevig2006, Moore2007,Fu2007, Konig2007,Hsieh2008} bloomed, the robustness of topological states has been at the forefront of modern condensed matter research~\cite{Hasan2010,Qi2011}.
As suggested by the bulk-boundary correspondence, the gapped bulk spectra of a two-dimensional (2D) TI are characterized by the Chern~\cite{Thouless1982,Klitzing1986,Haldane1988,Nagaosa2010} or the $\mathbb{Z}_2$~\cite{KaneII2005,FuII2006,Roy2009,Gresch2017,Soluyanov2011} invariant governed by the broken or protected time-reversal symmetry (TRS), while the boundaries on a nanoribbon geometry host robust conducting edge states.
One such prototype of a $\mathbb{Z}_2$ TI was proposed by Bernevig \etal~\cite{Bernevig2006}, which has been widely explored in explaining the robust helical QSH edges on a ribbon-like geometry corresponding to the inverted bulk bands of an HgTe/CdTe quantum well.
Meanwhile, the surge for generating nontrivial phases in TIs induced by different mechanisms, such as Floquet engineering~\cite{Eckardt2017,Oka2019,Rudner2020,Bao2022}, and more recently, the interaction effects~\cite{Rachel2018} has remained unabated. 
Particularly, to our relevance, the systems subject to periodic driving in time manifest nontrivial topology and edge modes in out-of-equilibrium scenarios~\cite{Eckardt2017,Oka2019,Rudner2020,Bao2022,Kitagawa2010,Kitagawa2011,Lindner2011,Gu2011,Rudner2013,Usaj2014} that are unattainable in static configurations. Interestingly, even if the static Hamiltonian lies in the topologically trivial phase, under the unitary time evolution, the Floquet operator exhibits localized eigenstates near the boundaries due to the nontrivial winding of the driven wave function, and consequently, the \textit{anomalous $\pi$ modes} emerge at the edges ~\cite{Kitagawa2010,Rudner2013,Usaj2014,Eckardt2017,Jiang2011,Perez2014,Yao2017,Seshadri2019,Ghosh2022}. 
Due to the time-translational symmetry of these dynamical edge modes the Floquet topological insulators (FTIs), being more active than ever, demonstrate novel properties, such as the Floquet time crystal~\cite{Else2016,Khemani2016}, dynamical localization~\cite{Kayanuma2008,Nag2014,DAlessio2014}, higher-order harmonics~\cite{Seshadri2019,Ghosh2022,GhoshI2024} etc.
With the advent of futuristic experimental tools, the distinct driven phases are realized in photonics~\cite{Rechtsman2013,Maczewsky2017}, acoustic~\cite{Peng2016,Fleury2016}, and atomic~\cite{Wintersperger2020} systems, further confirming the importance of this field.

We now switch our attention to the prime focus of this study, that is, the EPI-modulated scenario, which may give rise to richer topological phases in a QSH insulator (QSHI).
Although past encounters have evidenced the thermal aspects of EPI-induced topology~\cite{Garate2013,Li2013,Saha2014,Hu2021}, more recently, the role of coherent phonons has been established in the unveiling of distinct topological features~\cite{Calvo2018,Cangemi2019,Camacho2019,Farre2020,Pimenov2021,Medina2022,Lu2023,Shaozhi2023,Islam2024,Bhattacharyya2024,Min2025}. The interaction of a mobile electron with the oscillating lattice field produces \textit{\text{`dressed'}} quasiparticles, called polarons~\cite{Alexandrov2010}, which intrinsically modify the band topology accompanied by the gap-closing phenomena~\cite{Cangemi2019,Lu2023,Islam2024,Bhattacharyya2024}. topological 
Specifically, studies on phonon-induced topological signatures in hexagonal lattices~\cite{Calvo2018,Cangemi2019,Lu2023,Islam2024,Bhattacharyya2024}, chiral phonons~\cite{Hanyu2018,Medina2022}, Mott insulator~\cite{Farre2020}, topological superconductivity~\cite{Shaozhi2023}, polaronic Hall conductivity~\cite{Camacho2019,Pimenov2021}, etc., add further merit to EPI-based topological engineering in 2D materials. 
Nevertheless, dealing with such polaronic effects in the Floquet scenario may require additional care as the nontrivial winding of the state is now controlled by both the Floquet mechanism and EPI simultaneously, which results in limited exposure on the topic.
Although studies have advocated the phonon-induced Floquet features~\cite{Chaudhary2020,Hubener2018,Murakami2017}, the intricacies of handling the interplay between an FQSH phase and an EPI-driven TI phase with \textit{full} characterization have not been addressed to the best of our knowledge. 

To motivate our goal further, we emphasize the background of characterizing the FTI phases. While the first-order FTI phases are characterized through the conventional Zak phase, the Chern or the winding number~\cite{Asboth2014,Dal2015,Yao2017, Perez2015,Zhang2020,Shi2022} and Floquet $\mathbb{Z}_2$-invariant~\cite{Zhang2024, He2019, Schuster2019}, the higher-order FTI modes are designated either by mirror-graded winding number~\cite{Seshadri2019}, polarization~\cite{Ghosh2022} or the quadrupole moment~\cite{GhoshI2024}, depending upon the underlying symmetries of the system where both the zero and $\pi$ modes are gapped.
However, if a mixed-state situation arises (one of them being gapped and the other being gapless), the above indexing schemes fail, which requires one to rely on the \textit{spectral localizer} ($\mathcal{SL}$)~\cite{Loring2015,Loring2019,Loring2022,Cerjan2022}.
In contrast to conventional invariants, this renders an alternate characterization through the local and energy-resolved invariants in real-space, measuring whether the states at given energies can be \textit{continued} to the atomic limit (commutative space of the Hamiltonian and the position operator) with or without closing a bulk gap or altering any symmetry~\cite{Loring2019,Cerjan2022}. Especially, $\mathcal{SL}$ is employed in systems, such as the topological metals~\cite{Cerjan2022}, quasicrystals~\cite{Fulga2016,LoringII2019}, time-quasiperiodic Kitaev chains~\cite{Qi2024}, disordered semimetals~\cite{Franca2024,Schulz-Baldes2021}, driven Chern insulators~\cite{Liu2023,GhoshII2024}, nonlinear topological material~\cite{Wong2023}, photonics~\cite{CerjanII2022,Dixon2023} etc. where the conventional Altland-Zirnbauer scheme~\cite{Altland1997} fails.
However, its implementation in an interacting picture, especially in a QSHI, has not yet been reported, which intrigued us to attempt this study.  

To this end, the following pertinent questions naturally arise: 
Is it possible to generate emergent Floquet topological phases and transitions therein, solely by tuning the EPI strength?
Can a coexistence of gapped zero ($\pi$) and gapless $\pi$ (zero) modes emerge through EPI? If yes, can the $\mathcal{SL}$ prescription \textit{resolve} this gapless topology by characterizing it fully?
To address these questions, here we start with the Floquet analysis of the bare (non-interacting) BHZ model and briefly encapsulate its plausible FQSH phases (zero, $\pi$, and $0+\pi$) under suitable driving protocols before delving into the Floquet scenario of a \textit{dressed} (modified by EPI) BHZ model. We then show the generation of additional emergent FQSH phases solely via EPI and characterize the individual zero and $\pi$ phases by the conventional Chern number.
Upon extracting these phases, we arrive at a more intriguing situation where a gapless zero ($\pi$) and a gapped $\pi$ (zero) energy sectors coexist entirely due to EPI, thereby failing the conventional characterization. To characterize this distinct coexisting phase, we follow the local and energy-resolved $\mathcal{SL}$ prescription and compute a real-space Chern marker, maneuvering a successful characterization of all FQSH phases in our system.
These observations are further verified through the edge state analyses on a ribbon geometry, which explicitly corroborates the switching between these phases as the EPI strength is tuned.   

\textit{Floquet analysis of the bare BHZ Hamiltonian.} 
The static toy model of a 4-band BHZ Hamiltonian on a square lattice is usually described by the following states at each site: $|s,\uparrow\rangle$, $|s,\downarrow\rangle$, $|p_x + ip_y,\uparrow\rangle$ and $|p_x-ip_y,\downarrow\rangle$.
Under completely periodic boundary conditions, the BHZ Hamiltonian assumes a form in the $k$-space as~\cite{Bernevig2006}
\setlength\abovedisplayskip{3pt}%
\setlength\belowdisplayskip{3pt}%
\begin{align}
\mathcal{H}_{\text{BHZ}} = &\left[\mathcal{M}-2t(\cos k_x + \cos k_y)\right]\sigma_0\otimes\tau_z + \nonumber \\ &2t_{sp}\left[\sin k_x(\sigma_z\otimes\tau_x)+\sin k_y(\sigma_0\otimes\tau_y)\right].
\label{Ham:static BHZ}
\end{align}
Here, $\mathcal{M}$, $t$ and $t_{sp}$ correspond to the real-valued parameters of the BHZ Hamiltonian~\cite{Bernevig2006}, and $\sigma$ and $\tau$ represent the spin and orbital degrees of freedom, respectively.
While the static version of the BHZ Hamiltonian in addition to its nuances already exhibits rich topological phenomena, it is well known that a time-periodic drive applied to such a system augments its topological features.
Therefore, we begin our expedition with a driven version of the BHZ Hamiltonian, where we subject the system to the following step drive protocols:
\setlength\abovedisplayskip{1pt}%
\setlength\belowdisplayskip{1pt}%
\begin{align}
\mathcal{H} = &\mathcal{F}_1\mathcal{H}_1, \hspace{3mm}\text{for }~~0<t<\frac{T}{2},\nonumber\\
=&\mathcal{F}_2\mathcal{H}_2, \hspace{3mm}\text{for }~~\frac{T}{2}<t<T.
\label{Eq: drive}
\end{align}
Here, $\mathcal{H}_1$ ($\mathcal{H}_2$) corresponds to the  Hamiltonian for the first (second) half of the time period ($T$) multiplied by the amplitude $\mathcal{F}_1$ ($\mathcal{F}_2$) and is given by $\mathcal{H}_1=\mathcal{M}\sigma_0\otimes\tau_z$ containing only the onsite mass term, while $\mathcal{H}_2$ denotes the full static BHZ Hamiltonian~\eqref{Ham:static BHZ}.
Here, the time evolution operator can be evaluated via,
\setlength\abovedisplayskip{3pt}%
\setlength\belowdisplayskip{3pt}%
\begin{equation}
\mathcal{U}(T)\!=\!\exp{\biggl(\frac{-\text{i}\mathcal{F}_2\mathcal{H}_2}{2}\biggr)}\exp{\biggl(\frac{-\text{i}\mathcal{F}_1\mathcal{H}_1}{2}\biggr)}=\exp(-\text{i}\mathcal{H}_F),
\end{equation}
where we assume the value of $T$ to be equal to $1$.
\begin{figure}
\includegraphics[width=\columnwidth]{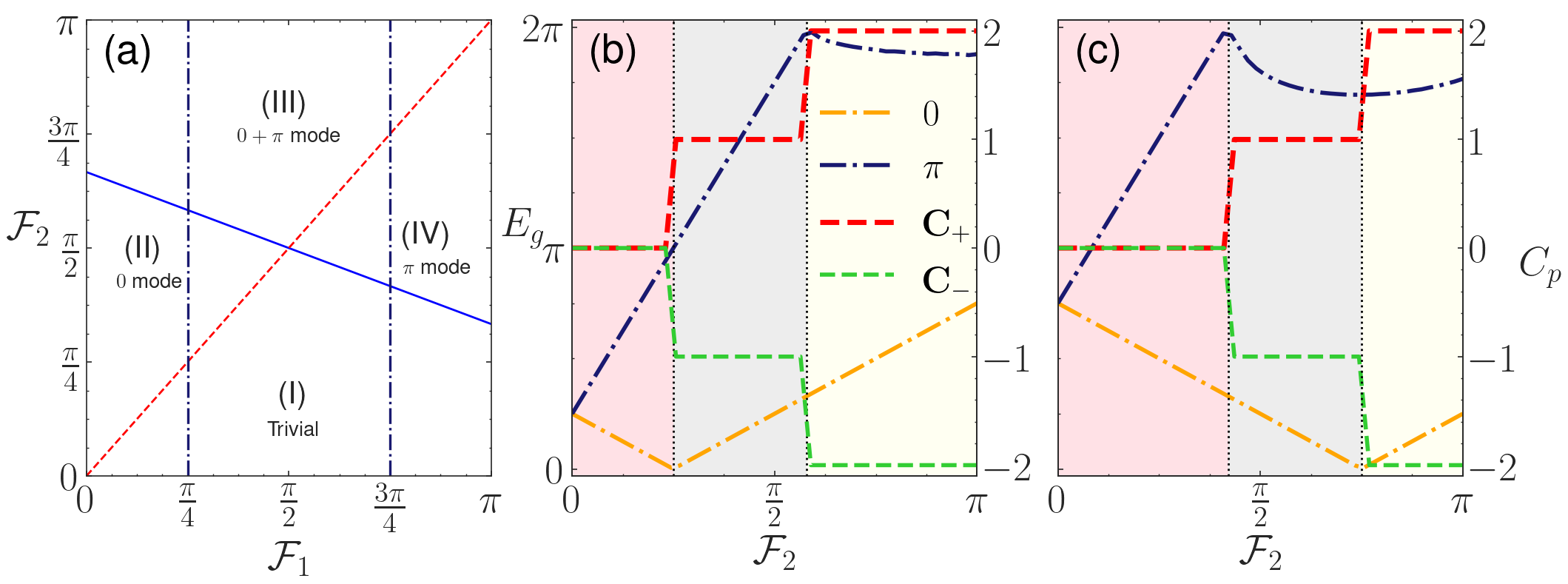}
\caption{(a) The phase diagram of the bare Floquet-BHZ model in the $\mathcal{F}_1$-$\mathcal{F}_2$ plane subjected to a step drive (Eq.~\ref{Eq: drive}) is shown. (b) The yellow (blue) dot-dashed line corresponds to the zero ($\pi$) gap-closing points as a function of $\mathcal{F}_2$ along $\mathcal{F}_1=\frac{\pi}{4}$. The variation of the PSCN as a function of $\mathcal{F}_2$ (shown by red and green dashed lines) shows a discontinuous change as soon as zero ($\pi$) $E_g$ line hits the value zero ($2\pi$). (c) A similar variation of the energy gap and PSCN as a function of $\mathcal{F}_2$ is shown for $\mathcal{F}_1=\frac{3\pi}{4}$.}
\label{fig: bare_floquet_phases}
\end{figure}
It is important to mention here that the eigenvalues $\phi_j$ of the Floquet evolution operator $\mathcal{U}(T)$ are the phases of a complex number of unit magnitude and are given by the following equation:
$\mathcal{U}|\psi_j\rangle=e^{\text{i}\phi_j}|\psi_j\rangle$, where $|\psi_j\rangle$ corresponds to the Floquet eigenstates.
This inherently implies the existence of two relevant gaps in the Floquet bandstructure, one being at zero energy and the other at $\pi$ energy, therefore enabling the existence of two variants of topological states at the aforementioned gaps.
The existence of such $\pi$ energy states clearly possesses no static analogue.
Numerical analysis of the time evolution operator $\mathcal{U}$ shows the existence of several zero and $\pi$ gap-closing transitions as a function of the amplitudes $\mathcal{F}_1$ and $\mathcal{F}_2$. 
A complete phase plot exhibiting the points of gap-closure on the $\mathcal{F}_1-\mathcal{F}_2$ plane is shown in Fig.~\ref{fig: bare_floquet_phases}(a).
Here, the dotted red line (solid blue line) corresponds to the points of zero ($\pi$) energy gap closure in the $\mathcal{F}_1-\mathcal{F}_2$ plane.
The phases marked (I), (II), (III) and (IV) correspond to regions of no edge (trivial), only zero, both zero and $\pi$, and only $\pi$ energy edge states respectively.
For concreteness, we first fix the value of $\mathcal{F}_1$ to $\frac{\pi}{4}$ and then to $\frac{3\pi}{4}$ to explicitly show the points of zero and $\pi$ energy gap closure as a function of the parameter $\mathcal{F}_2$, as shown by the black dot-dashed lines in Fig. \ref{fig: bare_floquet_phases}(a).
To characterize the topology of the various phases obtained by the gap-closing transition, we plot the variation of the projected spin Chern number (PSCN) $C_p$ obtained by integrating the projected Berry curvature $\Omega_p(\vec k)\!=\!i\biggl[\! \biggl<\!\frac{\partial \psi_p(\vec k)}{\partial k_x}\!\biggl|\!\frac{\partial \psi_p(\vec k)}{\partial k_y}\!\biggl> - \biggl <\!\frac{\partial \psi_p(\vec k)}{\partial k_y}\!\biggl|\!\frac{\partial \psi_p(\vec k)}{\partial k_x}\!\biggl>\!\biggr]$
over the entire Brillouin Zone. It is to be noted here that the two-fold degenerate eigenstates of the occupied subspace are extracted using the projector $P=(\mathbb{I}\pm S)/2$, where $S=\sigma_z\otimes\sigma_0$, corresponds to a symmetry of the Floquet Hamiltonian $\mathcal{H}_F$.
These degenerate eigenstates have eigenvalues $\pm1$ when acted upon by the symmetry operator $S$, thereby enabling their extraction separately using the projector $P$.
We refer to these extracted states and their corresponding Chern numbers as $|\pm\rangle$ and $C_\pm$ respectively. 
The variation of $C_p$, where $p\in\pm$, clearly exhibits the topological phase transitions caused by the energy gap-closing in the system as a function of $\mathcal{F}_2$ (Figs.~\ref{fig: bare_floquet_phases}(b)-(c)).
The yellow (blue) dot-dashed line in Fig. \ref{fig: bare_floquet_phases}(b), (c) corresponds to the points of zero ($\pi$) energy gap closure as a function of $\mathcal{F}_2$ for $\mathcal{F}_1=\frac{\pi}{4}$ and $\frac{3\pi}{4}$ respectively.
The nontrivial Floquet topology manifests zero and $\pi$ edge states (as well as the coexistence of both) on a nanoribbon geometry.
These results set the premise for examining the EPI effects in the subsequent sections.

\textit{Dressed BHZ model.} 
To begin with, the intrinsic modification to the band topology can be explored via the interaction between the mobile charge carrier and the inherent lattice modes (unlike considering the phonon excitation as an external drive in Ref.~\cite{Murakami2017}) in the static limit to obtain a coherent polaronic state by canonically transforming the basis to a displaced oscillator ground state~\cite{Lang1963}. 
Secondly, prior to the application of the driving protocol, this resultant coherent state is assumed to provide an effective electronic model dressed by the phonon modes with all the symmetries of the non-interacting picture being intact as before and captures the (small) polaron physics~\cite{Holstein1959} in our system as well.   
The modified BHZ Hamiltonian including the Holstein-type EPI~\cite{Cangemi2019,Lu2023,Islam2024,Bhattacharyya2024} can be modeled as a tight-binding version of Eq.~\eqref{Ham:static BHZ} in a two-component orbital basis $c^\dagger=(c_{s,\sigma}^\dagger,c_{p,\sigma}^\dagger)$  as
\begin{figure*}
\begin{center}
\includegraphics[width=16cm]{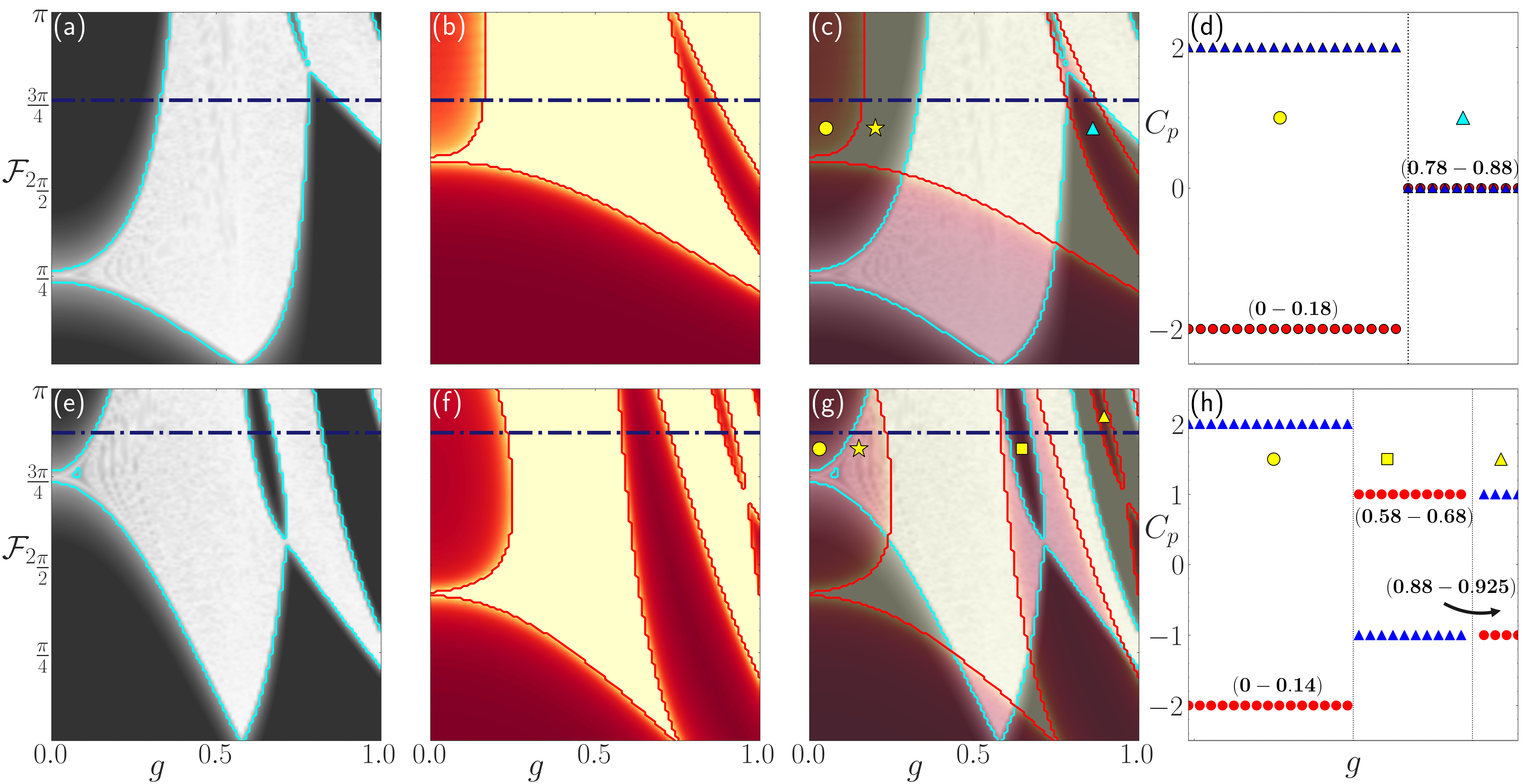} 
\caption{Upper Panel: The depiction of the (a) zero energy and (b) $\pi$ energy gap of the effective time evolution operator $U^\text{eff}(T)$ is shown in the $\mathcal{F}_2-g$ plane. The dark areas demarcated by the cyan (red) contour lines in (a) ((b)) correspond to the presence of a finite zero ($\pi$) energy gap in the spectrum for $\mathcal{F}_1=\frac{\pi}{4}$. (c) A universal plot formed by superimposing the band gaps in (a) and (b) is displayed. The blue dot-dashed line at $\mathcal{F}_2=\frac{3\pi}{4}$, captures the maximum number of gap-closing transitions as a function of $g$ that result in distinct phases which are marked by  \protect\filledcircle{yellow}, \protect\mystaroutline,~ and \protect\filledtriangle{cyan!35} symbols. (d) The PSCN, $C_p$ (labeled by \protect\filledtriangle{blue} and \protect\filledcircle{red} for the $|+\rangle$ and $|-\rangle$ states, respectively) plotted as a function of $g$ shows an absolute value of $2$ for the phase marked \protect\filledcircle{yellow} and $0$ for the phase marked \protect\filledtriangle{cyan!35}. Lower panel: Similar to the upper panel (e) and (f) correspond to the zero and $\pi$ energy gap, respectively for $\mathcal{F}_1=\frac{3\pi}{4}$. (g) The universal plot superimposed by (e) and (f) is obtained, where the blue dot-dashed line is drawn at $\mathcal{F}_2=\frac{7\pi}{8}$. The phases of interest are marked by the \protect\filledcircle{yellow}, \protect\mystaroutline~, \protect\myfilledsquare{yellow}, and \protect\filledtriangle{yellow} symbols. (h) $C_p$ shows an absolute value of $2$ for \protect\filledcircle{yellow}, $1$ for \protect\myfilledsquare{yellow}, and $0$ for \protect\filledtriangle{yellow} marked phases. The \protect\mystaroutline-marked phases are characterized through $\mathcal{SL}$ in Fig.~\ref{fig: SL}.}

\label{fig: epi_floquet_phases}
\end{center}
\end{figure*}
\setlength\abovedisplayskip{3pt}%
\setlength\belowdisplayskip{3pt}%
\begin{eqnarray}
 \mathcal{H}_{\uparrow,\downarrow}\!&=&\!\sum_{i,j}[c^{\dagger}_{i,j}\mathcal{M}\tau_z c_{i,j}\!+\!tc^{\dagger}_{i,j}\tau_z(\!c_{i+1,j}\!+\!c_{i,j+1})\nonumber\\
&&\mp\!it_{sp}(\!c^{\dagger}_{i,j}\tau_xc_{i+1,j}\!+c^{\dagger}_{i,j}\tau_yc_{i,j+1})]\nonumber\\
&&+\hbar\omega_{0}\!\sum_{r\in(i,j)}\biggl[\biggl(b^{\dagger}_{r}b_{r}\!+\!\frac{1}{2}\biggr)+gc^{\dagger}_{r}c_{r}(b^{\dagger}_{r}+b_{r})\biggr]I_2,
\label{Ham:pol_BHZ}
\end{eqnarray}
where the last two terms containing the phonon operators $(b^\dagger,b)$ stem from the phononic contributions. While the former represents the total energy of the longitudinal optical (LO) phonons vibrating with a dispersionless frequency $\omega_0$, the latter describes an onsite coupling ($g$ being the coupling strength) of the LO phonons with the local electronic densities at the same site which produces strong lattice polaronic effects. In the anti-adiabatic (the LO phonons being much faster smoothly follow  the electronic motion without changing their distributions), i.e., the high-frequency ($\hbar\omega_0\!>>\!t,t_{sp},g$) limit, these phonon modes are decoupled by applying the coherent Lang-Firsov transformation~\cite{Lang1963}: $\tilde{{\mathcal{H}}}=e^\mathcal{R}\mathcal{H}e^{-\mathcal{R}}$, $\mathcal{R}$ being the generator given by $\mathcal{R}\!=\!g\sum_r c^{\dagger}_{r}c_{r}(b^{\dagger}_{r}-b_{r})$,  which yields an effective electronic model (dressed by phonon clouds) via a zero-phonon averaging: $\langle 0|\tilde{\mathcal{H}}|0\rangle=\mathcal{H}^{\text{eff}}$ as
\setlength\abovedisplayskip{3pt}%
\setlength\belowdisplayskip{3pt}%
\begin{align}
 \mathcal{H}_{\uparrow}^{\text{eff}}=&\sum_{i,j}\biggl[c^{\dagger}_{i,j}\tilde{\mathcal{E}} c_{i,j}+c^{\dagger}_{i,j}\tilde{T}_xc_{i+1,j}+c^{\dagger}_{i,j}\tilde{T}_yc_{i,j+1}\biggr], \nonumber\\
\mathcal{H}_{\downarrow}^{\text{eff}}=&\mathcal{H}_{\uparrow}^{\text{eff}}:\tilde{T}\rightarrow\tilde{T}^*,
\label{Ham:dressed BHZ}
\end{align}
where the renormalized parameters
$\tilde{\mathcal{E}}\!=\!\mathcal{M}\tau_z\!-\!g^2\hbar\omega_0I_2$, $\tilde{T}_x(\tilde{T}_y)\!=\!\tilde{t}\tau_z\!-\!i\tilde{t}_{sp}\tau_x(\tau_y)$ and $\tilde{t}(\tilde{t}_{sp})\!=\!t(t_{sp})e^{-g^2}$ infuse the polaronic contributions through $g$ which significantly modify the bulk bands and hence the band topology because of a band narrowing effect caused by the Holstein reduction factor ($e^{-g^2}$) and the polaronic self-shift energy ($-g^2\hbar\omega_0$). Hence, the Floquet version of Eq.~\eqref{Ham:dressed BHZ} is referred to as the \text{`dressed Floquet-BHZ'} model for which the time evolution operator is now given as
\setlength\abovedisplayskip{3pt}%
\setlength\belowdisplayskip{3pt}%
\begin{equation}
\mathcal{U}^{\text{eff}}(T)\!=\!\exp{\biggl(\frac{-\text{i}\mathcal{F}_2\mathcal{H}_2^{\text{eff}}}{2}\biggr)}\exp{\biggl(\frac{-\text{i}\mathcal{F}_1\mathcal{H}_1^{\text{eff}}}{2}\biggr)}=e^{-\text{i}\mathcal{H}_F^{\text{eff}}},
\end{equation}
where $\mathcal{H}_1^{\text{eff}}$ ($\mathcal{H}_2^{\text{eff}}$) corresponds to the effective Hamiltonian with renormalized parameters ($\tilde{\mathcal{E}}$, $\tilde{t}$, $\tilde{t}_{sp}$) for the first (second) half of the time period $T$, thereby representing the effective Floquet Hamiltonian as $\mathcal{H}_F^\text{eff}$.
The effective time evolution operator $\mathcal{U}^\text{eff}(T)$ is now numerically analyzed in our subsequent study.

\textit{EPI-induced emergent FQSH phases and their topological characterizations.} To ascertain the enhancement in topological features of the dressed Floquet-BHZ model caused by the inclusion of EPI, we first consider the non-interacting scenario of Fig. \ref{fig: bare_floquet_phases}(b), that is $\mathcal{F}_1=\frac{\pi}{4}$ and plot the zero and $\pi$ energy gap of the interacting Floquet-BHZ system in the $\mathcal{F}_2-g$ plane (Fig. \ref{fig: epi_floquet_phases}(a), (b) respectively).
In Fig. \ref{fig: epi_floquet_phases}(a) (Fig. \ref{fig: epi_floquet_phases}(b)), the dark regions bordered by the cyan (red) contours correspond to the areas with a finite zero ($\pi$) energy gap.
Furthermore, we superimpose the phase diagrams for the zero and $\pi$ energy gap (for $\mathcal{F}_1=\frac{\pi}{4}$) in a universal phase plot to clearly depict the Floquet phase transitions carried out by the interplay of $\mathcal{F}_2$ and EPI strength $g$ (Fig. \ref{fig: epi_floquet_phases}(c)).
We observe the possibility of multiple phase transitions caused by several energy gap-closings purely as a function of the EPI strength $g$ for a fixed value of $\mathcal{F}_2$, thereby enabling the extraction of distinct Floquet topological phases, controlled by the parameter $g$ solely.
To explore this possibility further, we fix the value of $\mathcal{F}_2$ to $\frac{3\pi}{4}$ in Fig. \ref{fig: epi_floquet_phases}(c) (so that maximum number of nontrivial phases is encompassed) and characterize them using suitable topological invariants for several values of $g$ within the gapped regions.
It can be clearly seen that the system sweeps through three distinct topological phases as a function of $g$ for $(\mathcal{F}_1,\mathcal{F}_2)=(\frac{\pi}{4},\frac{3\pi}{4})$.
These regions include the ones designated by a yellow circle (for $0<g<0.18$ where both zero and $\pi$ energy gaps are open), a yellow star (for $0.18<g<0.33$ where the zero energy gap is open and the $\pi$ energy gap is closed) and a cyan triangle (for $0.78<g<0.88$ where both zero and $\pi$ energy gaps are again open). 
To determine the topology of the bulk for the aforementioned phases, we now resort to the evaluation of the PSCN, $C_p$.
We observe that $|C_p|$ assumes a value of $2$ for $0<g<0.18$ as marked in Fig.~\ref{fig: epi_floquet_phases}(c) by the yellow circle.
This hints at the existence of both zero and $\pi$ energy edge states for this regime of EPI strength.
Furthermore, for the phase marked by the cyan triangle in Fig.~\ref{fig: epi_floquet_phases}(c) ($0.78<g<0.88$), $|C_p|$ becomes zero, indicating a purely trivial insulating phase.
We now move to explore similar variations in the $\mathcal{F}_2-g$ plane for $\mathcal{F}_1=\frac{3\pi}{4}$ (Fig.~\ref{fig: epi_floquet_phases}(e), (f) and (g)).
Again, multiple phase transitions are seen to occur as a function of the EPI and we fix the value of $\mathcal{F}_2$ at $\frac{7\pi}{8}$ to encounter as many nontrivial phase transitions as possible.
The different topological phases in the universal phase plot (Fig.~\ref{fig: epi_floquet_phases}(g)) for $(\mathcal{F}_1,\mathcal{F}_2)=(\frac{3\pi}{4},\frac{7\pi}{8})$ are marked by a yellow circle (for $0<g<0.14$ where both zero and $\pi$ energy gaps are open), a yellow star (for $0.14<g<0.24$ where the zero energy gap is closed but the $\pi$ energy is open), a yellow square (for $0.58<g<0.68$ where only zero energy gap is open) and a yellow triangle (for $0.88<g<0.93$ where only $\pi$ energy gap is open). 
On evaluation of the PSCN, $|C_p|$, shows a value of $2$ for the phase marked by the yellow circle and $1$ for both the phases marked by the yellow square and the yellow triangle.
This implies the existence of both zero and $\pi$ energy edge modes for the former phase and either zero or $\pi$ energy edge modes for the latter two phases. Notably, the emergence of these phases is a direct consequence of a nontrivial winding of the Floquet bands mediated via an EPI where the Holstein factors in Eq.~\eqref{Ham:dressed BHZ} account for the polaron-driven band topology.
The transitions in $|C_p|$ with respect to $g$ (Figs.~\ref{fig: epi_floquet_phases}(d) and (h)) further suggest the change in the Hall conductivity responses at the edges solely through EPI that indicates phonon-induced topological transport.
\begin{figure}
\begin{center}
\includegraphics[width=\columnwidth]{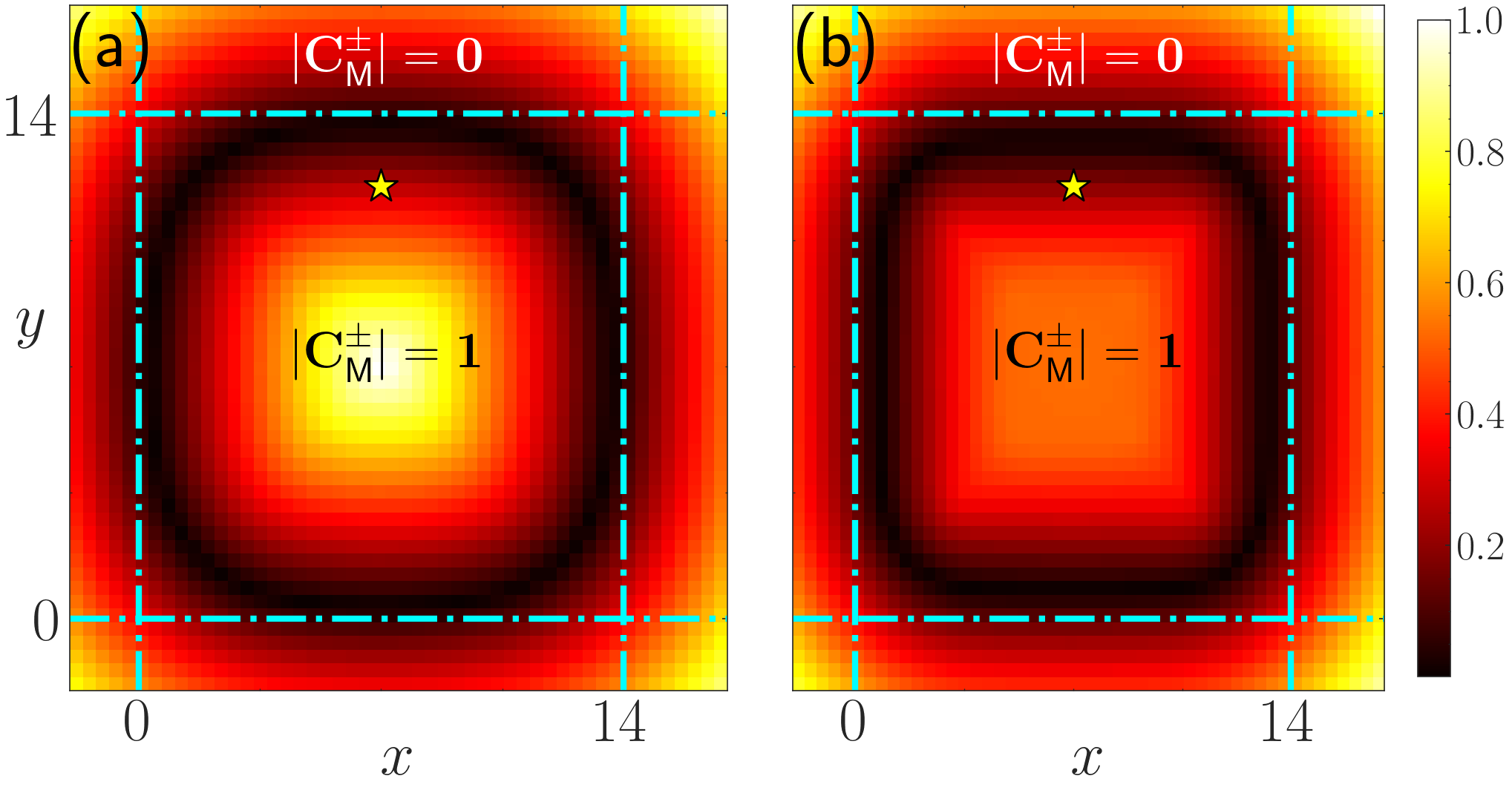}
\caption{Topological characterization of the (coexistent gapped and gapless) phase marked by \protect\mystaroutline~in Figs.~\ref{fig: epi_floquet_phases}(c) and (g) is depicted through the localizer gap for (a) $(\mathcal{F}_1, \mathcal{F}_2)=(\frac{\pi}{4}, \frac{3\pi}{4})$ and (b) $(\mathcal{F}_1, \mathcal{F}_2)=(\frac{3\pi}{4}, \frac{7\pi}{8})$ at $g=0.2$. The closure of the normalized localizer gap at the boundaries of the square supercell (marked by the dot-dashed lines) indicates a topological zero energy bulk gap for (a) and a topological $\pi$ energy bulk gap for (b). The Chern markers shown in the figure provide additional information pertaining to the topology of the system and are evaluated to be equal to $\pm1$ ($0$) for the topological interior (trivial exterior).}
\label{fig: SL}
\end{center}
\end{figure}

We now come to the highlight of our analysis, where the EPI strength $g$ leads to the formation of a phase, marked by the yellow star in Figs.~\ref{fig: epi_floquet_phases}(c) and (g) exhibiting the coexistence of a gapless $\pi$ (zero) and gapped zero ($\pi$) energy bulk.
This presents a situation where the prescription for the evaluation of conventional bulk topological invariants fails owing to the absence of a continuous and distinct bulk gap.
Therefore, the Chern number, which can characterize the topology of the bulk gap for the phases previously described (having both the zero and the $\pi$ energy gap open), can no longer be used now. 
Instead, we resort to evaluating the $\mathcal{SL}$ to decipher the nature of the zero or $\pi$-energy gap, while the other energy sector is gapless within this regime of $g$.
For a 2D system, the $\mathcal{SL}$ is defined as~\cite{Loring2015,Loring2019,Loring2022,Cerjan2022}
\begin{align}
\mathcal{L}=\kappa[(X-xI)\sigma_x + (Y-yI)\sigma_y] + (\mathcal{H}-EI)\sigma_z
\end{align}
where $X$ ($Y$) corresponds to the position operators for a finite supercell and $\mathcal{H}$ denotes the real-space Hamiltonian describing the system.
Furthermore, the parameter $\kappa$ ($>0$) serves to ensure dimensional compatibility between the Hamiltonian and the position operators. 
Evidently, the $\mathcal{SL}$ represents a real-space description of topological nontriviality, determining whether the system hosts an eigenstate with energy in the vicinity of $E$, which is also approximately localized at the positions denoted by the coordinates $x$ and $y$.
Interestingly, $x$, $y$, and $E$ can assume values outside the system's physical boundaries and energy spectrum, respectively.
Given that the $\mathcal{SL}$ possesses a spectrum denoted by $\sigma_{\mathcal{L}}^{x, y, E}(X, Y, \mathcal{H})$, the gap of the localizer spectrum, which is \textbf{min}[$|\sigma_{\mathcal{L}}^{x, y, E}(X, Y, \mathcal{H})|$], determines the existence (or absence) of topological boundary states.
In fact, the closure of the localizer gap at a given value of $x$, $y$ and $E$, indicates the presence of an eigenstate of the system close to those given parameters in the position and spectral space.
This in turn acts as a convenient indicator of topology in Floquet systems which implies that the existence of topological boundary states at the zero or $\pi$ energy gap of a Floquet Hamiltonian, would require the localizer gap to be closed for $x$ and $y$ values pertaining to the boundary of the corresponding supercell at $E=0 \text{ or }\pi$.
We therefore, diagonalize the $\mathcal{SL}$ of the Hamiltonian $\mathcal{H}^\text{eff}_F$ on a range of $x$ and $y$ values ($-2<x, y<12$) for $E=0$ ($E=\pi$) for the starred phase in Fig. \ref{fig: epi_floquet_phases}(c) ((g)).
This is done to make sure that the boundaries of the supercell lie well within the range chosen for $x$ and $y$.
The gap of the $\mathcal{SL}$, that is, $\sigma^{\textbf{min}}_{\mathcal{L}}=$\textbf{min}[$|\sigma_{\mathcal{L}}^{x, y, E}(X, Y, \mathcal{H})|$], should show a closing transition at the boundary of the system for $E=0$ and $E=\pi$ for the starred phases in Fig. \ref{fig: epi_floquet_phases}(c) and (g) respectively in case the relevant gaps in the energy spectrum are topological.
The normalized localizer gap, that is $\sigma_\text{min}^\mathcal{L}/\textbf{max}[\sigma_\text{min}^\mathcal{L}]$ as shown in Fig. \ref{fig: SL}(a) and (b) indeed show such a transition, implying that the zero ($\pi$) energy gap of the aforementioned phase in Fig. \ref{fig: epi_floquet_phases}(c) ((g)) is topologically nontrivial.
This predicts the presence of topological boundary state within the zero ($\pi$) energy gap in Fig. \ref{fig: SL}(a) (Fig. \ref{fig: SL}(b)) in coexistence with a gapless $\pi$ (zero) energy bulk.
It is important to mention here that, our system being modeled by a 4-band effective Floquet Hamiltonian ($\mathcal{H}^{\text{eff}}_F$), we can employ the symmetry operator $S$ to block-diagonalize it in real-space ($\mathcal{H}^{\text{eff}, \pm}_F$) and apply the $\mathcal{SL}$ ($\mathcal{L}^{\pm}$) on the projected up and down spin blocks separately.
This is done such that a Chern marker ($C_{M}^{\pm}$) corresponding to $\mathcal{L^\pm}$ for the blocks of $\mathcal{H}_F^{\text{eff}}$ (i.e., $\mathcal{H}_F^{\text{eff},\pm}$), can be defined as
\begin{equation}
C_{M}^{\pm}=\frac{1}{2}\text{sig }\mathcal{L}(X, Y, \mathcal{H}_F^{\text{eff},\pm})=\frac{1}{2}\text{sig }\mathcal{L}^\pm,
\end{equation}
where sig corresponds to the difference in positive and negative eigenvalues of the $\mathcal{L}^\pm$.
The Chern markers inside and outside the extent of the system have been marked in Figs.~\ref{fig: SL}(a) and (b) which clearly shows that within the bounds of the supercell for both (a) and (b), the topological index is nontrivial ($|C_{M}^{\pm}|=1$), whereas it is trivial ($C_{M}^{\pm}=0$) outside. Thus, the change in these topological markers corroborates topological phase transitions induced solely by EPI. We summarize our results in Table.~\ref{table}.
\begin{table}[h!]
    \centering
    \begin{tabular}{
  >{\centering\arraybackslash}p{0.4cm}
  >{\centering\arraybackslash}p{1.15cm}|
  >{\centering\arraybackslash}p{1.25cm}
  >{\centering\arraybackslash}p{1.75cm}
  >{\centering\arraybackslash}p{1.65cm}
  >{\centering\arraybackslash}p{1.65cm}
} 
\hline\hline  
& Phases
& $\filledcircle{yellow}$ 
& $\mystaroutline$ 
& $\myfilledsquare{yellow}$ 
& $\filledtriangle{cyan!35}$/$\filledtriangle{yellow}$\\
\hline
\rotatebox[origin=c]{90}{$(\mathcal{F}_1,\mathcal{F}_2)\!=\!(\frac{\pi}{4}\!,\!\frac{3\pi}{4})$}  & \begin{tabular}{@{}c@{}}Modes\\\\\\Chern\\marker\\\\$g$-range\end{tabular} 
& \begin{tabular}{@{}c@{}}Zero\\and $\pi$\\\\\\$C_p=\pm2$\\\\$[0:0.18]$\end{tabular} 
& \begin{tabular}{@{}c@{}}Gapped zero, \\gapless $\pi$\\(coexistent)\\\\ $C_{M}^{\pm}=\pm 1$\\\\$[0.18:0.33]$\end{tabular} 
& \begin{tabular}{@{}c@{}}\\\\\end{tabular} 
& \begin{tabular}{@{}c@{}}Trivial ($\filledtriangle{cyan!35}$)\\\\\\\\$C_p=0$\\\\$[0.78:0.88]$\end{tabular} \\
\hline
\rotatebox[origin=c]{90}{$(\mathcal{F}_1,\mathcal{F}_2)\!=\!(\frac{3\pi}{4}\!,\!\frac{7\pi}{8})$} & \begin{tabular}{@{}c@{}}Modes\\\\\\Chern \\marker\\\\$g$-range\end{tabular} 
& \begin{tabular}{@{}c@{}}Zero\\and $\pi$\\\\\\$C_p=\pm2$\\\\$[0:0.14]$\end{tabular} 
& \begin{tabular}{@{}c@{}}Gapped $\pi$, \\gapless zero\\(coexistent)\\\\ $C_{M}^{\pm}=\pm 1$\\\\$[0.14:0.24]$\end{tabular} 
& \begin{tabular}{@{}c@{}}Only\\ zero\\\\\\$C_p=\pm1$\\\\$[0.58:0.68]$\end{tabular} 
& \begin{tabular}{@{}c@{}}Only $\pi$\\ ($\filledtriangle{yellow}$)\\\\\\$C_p=\pm1$\\\\$[0.88:0.93]$\end{tabular} \\
\hline\hline
    \end{tabular}
    \caption{Table summarizing the results of the EPI-induced FQSH phases obtained for our system with their characterization for different combinations of $\mathcal{F}_1$ and $\mathcal{F}_2$. The findings can be correlated with Fig. \ref{fig: epi_floquet_phases} and \ref{fig: SL}.}
    \label{table}
\end{table}

\begin{figure}
\includegraphics[width=\columnwidth]{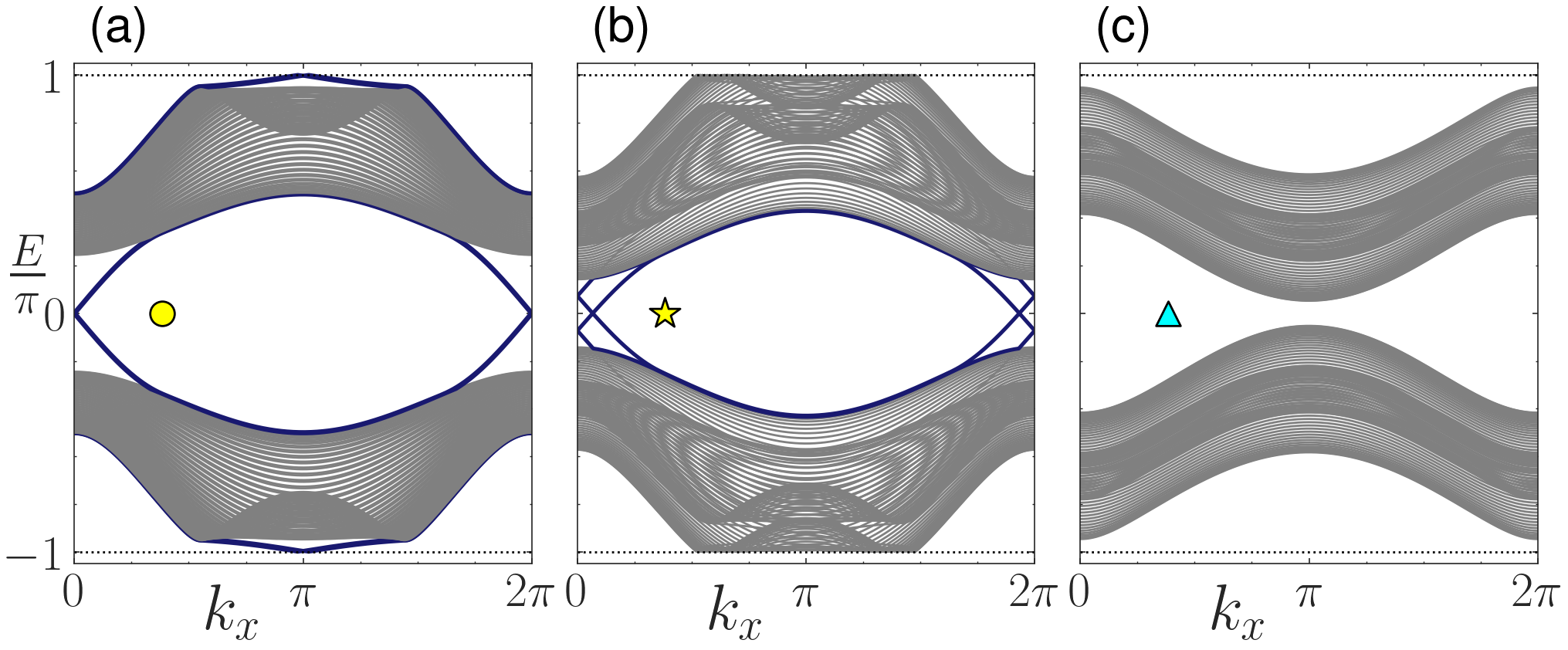}
\caption{Edge state spectra of a semi-infinite ribbon for $(\mathcal{F}_1,\mathcal{F}_2)=(\frac{\pi}{4},\frac{3\pi}{4})$ are shown as a function of $k_x$ for the phases marked by (a) \protect\filledcircle{yellow} (b) \protect\mystaroutline~, and (c) \protect\filledtriangle{cyan!35}. The appearance and vanishing of the helical edge states corresponding to their topological markers are summarized in Table.~\ref{table}.}
\label{fig: edge_F2=3pi/4}
\end{figure}

\begin{figure}
\includegraphics[height=8cm, width=0.8\columnwidth]{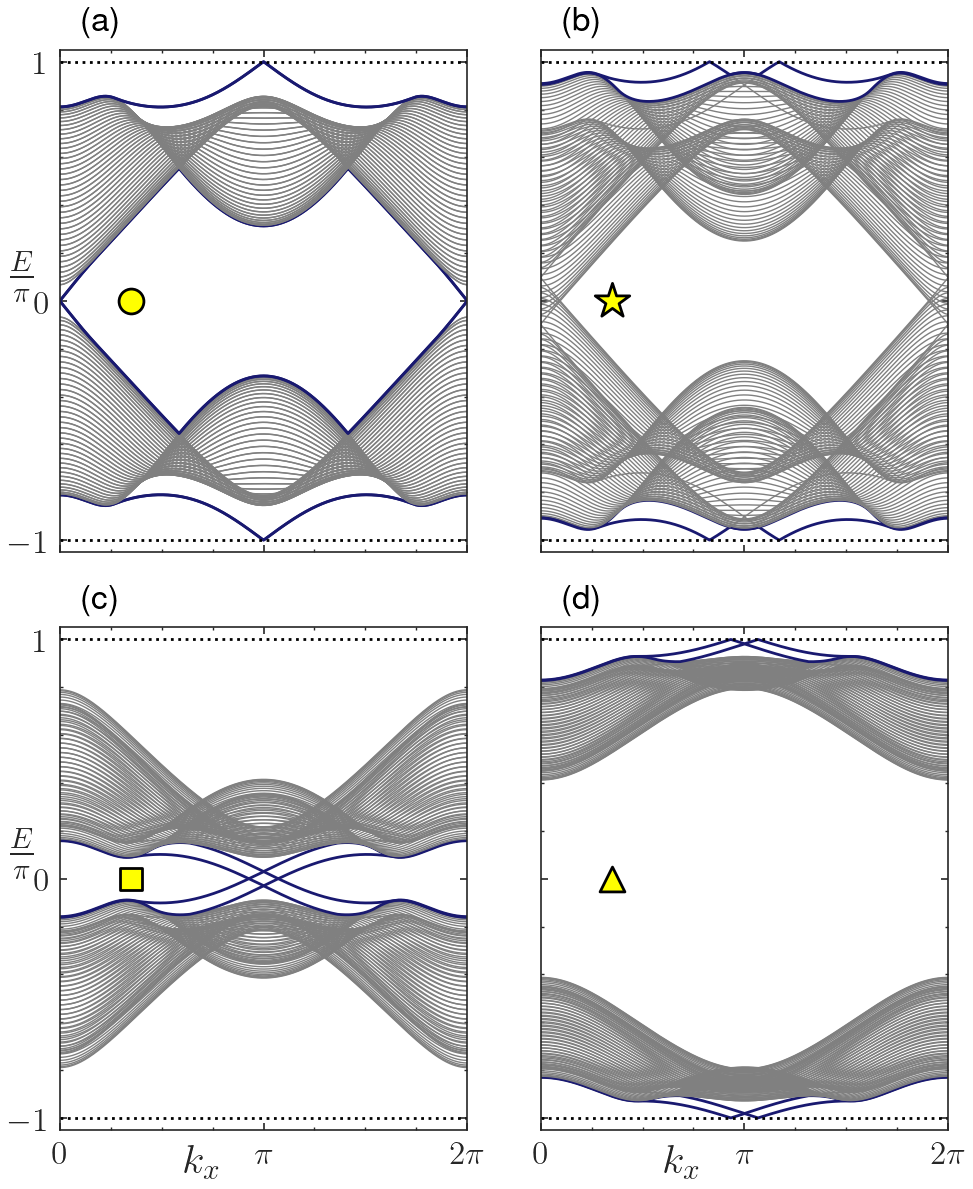}
\caption{Similar to Fig.~\ref{fig: edge_F2=3pi/4}, the edge state spectra are shown for $(\mathcal{F}_1,\mathcal{F}_2)=(\frac{3\pi}{4},\frac{7\pi}{8})$ representing the phases marked by (a) \protect\filledcircle{yellow}~ (b) \protect\mystaroutline, (c) \protect\myfilledsquare{yellow}~, and (d) \protect\filledtriangle{yellow}. The appearance of the helical edge states corresponding to their topological markers are summarized in Table.~\ref{table}.}
\label{fig: edge_F2=7pi/8}
\end{figure}

\textit{Edge state response to EPI.}~ In Figs.~\ref{fig: edge_F2=3pi/4} and \ref{fig: edge_F2=7pi/8}, we present the edge state characteristics of the dressed Floquet-BHZ model for different regimes of $g$ at which the FQSH phases are obtained in Figs.~\ref{fig: epi_floquet_phases}(c) and (g).
The edge states are computed on a semi-infinite ribbon-like geometry, periodic along the $x$-direction ($k_x$ being the corresponding Fourier conjugate momentum) and aperiodic along the $y$-direction.
As suggested by the Chern markers in Fig.~\ref{fig: epi_floquet_phases}(d), the distinct \text{`$0+\pi$'} (for $0< g< 0.18$) and \text{`no mode'} (for $0.78< g< 0.88$) phases marked by the yellow circle and the cyan triangle, respectively, host localized zero and $\pi$-energy helical edge states (Fig.~\ref{fig: edge_F2=3pi/4}(a)) at the boundaries for the former (corresponding to $|C_p|=2$) and no edge (trivial) states (Fig.~\ref{fig: edge_F2=3pi/4}(c)) for the latter phase (corresponding to $C_p=0$) along the $(\mathcal{F}_1,\mathcal{F}_2)=(\frac{\pi}{4},\frac{3\pi}{4})$ line.
More interestingly, for the mixed state (marked by yellow star), we notice clear evidence of the zero mode QSH helical edge states (Fig.~\ref{fig: edge_F2=3pi/4}(b)), while those of the $\pi$ mode merge into the bulk, confirming the coexistence of a gapped zero and a gapless $\pi$ phase caused solely due to $g$ ($0.18<g<0.33$).
This distinct phenomenon is also verified through the real-space $\mathcal{SL}$ characterization with a Chern marker $C_M^{\pm}=\pm 1$ and zero for the gapped topological interior and trivial exterior, respectively, in Fig.~\ref{fig: SL}(a). 
Similar observations also hold for the FQSH phases for $(\mathcal{F}_1,\mathcal{F}_2)=(\frac{3\pi}{4},\frac{7\pi}{8})$, shown in Figs.~\ref{fig: edge_F2=7pi/8}(a)-(d).
As $g$ is tuned, these figures respectively demonstrate the presence of both the helical zero and $\pi$ energy edge modes (yellow circle), coexistent mixed states (marked by a star) with the gapped $\pi$ mode edge states and gapless zero mode, only zero mode (yellow square), and only $\pi$ mode (yellow triangle).
These findings are inevitably guaranteed by the topological characterizations through the conventional Chern numbers and the $\mathcal{SL}$ prescription obtained in Fig.~\ref{fig: epi_floquet_phases}(h) and Fig.~\ref{fig: SL}(b), respectively.
Thus, the edge state signatures determine the probing of different FQSH phases and topological transitions between them upon tuning the EPI strength $g$.

\textit{Summary and conclusion.}
To summarize, we study the effect of EPI on a Floquet-BHZ model, subjected to a step drive. 
Remarkably, we observe that the inclusion of EPI leads to topological augmentation via forming a phonon-dressed coherent polaronic state that significantly modifies the band topology.
The manifestation of EPI in the topology of the Floquet-BHZ model is two-fold, the first being the smooth transition between the Floquet topological phases solely as a function of the EPI strength $g$ and the second being the coexistence of topologically gapped zero and gapless $\pi$ energy sectors or vice versa.
We provide a full characterization of all the emergent FQSH phases in our study, where a projected spin Chern number characterizes all the phases that have a distinct and continuous bulk gap.
However, such a bulk characterization fails when gapped and gapless phases coexist.
We thereby switch to the evaluation of a real-space indicator of topology, namely the spectral localizer, to characterize the coexisting emergent phases.
Furthermore, the edge behavior of all the topological phases generated by the EPI strength $g$ has been shown on a nanoribbon configuration and they perfectly agree with the characterization made beforehand.

We now briefly present some possible experimental perspectives that may aid in realizing our results. 
Optical lattices facilitate laser trapping of atoms, which mimic intricate solid-state systems with precise control over the parameters of the Hamiltonian~\cite{McKay2011, Wintersperger2020, Scheurer2015}, manifesting fruitful demonstration of photon-induced Floquet modes~\cite{Jiang2011,Eckardt2017,Oka2019,Wintersperger2020,Rudner2020,Giovannini2020, Bao2022}. Notably, the topological properties of a QSH insulator on a BHZ model have also been realized in a cold atom experiment~\cite{Lv2021}, which is of significant interest to the topological community.
The importance of optical lattices in achieving the Floquet phases being noted, the tunable EPI scenario, such as ours, can be conceptualized via trapping the polar molecules (such as CdTe that favors a polaronic state dressed by phonons) in an optical lattice~\cite{Bruderer2007,Herrera2011} to incorporate polaron-induced Floquet signatures in such a cold atom setup where strain~\cite{Si2016} or surface~\cite{Hwang2013,Aleshkin2024} engineering may influence phonon modes and the EPI strength can be modulated by an electric field~\cite{Yan2007}.  
Thus, aided by a robust handle over the parameters of the system, these setups in turn provide an ideal platform for realizing a step-driven and EPI-induced Floquet phases where the ability to tune the parameters precisely is of prime importance.

\textit{Acknowledgments.} S.L. acknowledges financial support from the Ministry of Education (MOE), Govt. of India, through the Prime Minister’s Research Fellowship (PMRF) scheme in May 2022. K.B. acknowledges financial support from the Anusandhan National Research Foundation (ANRF), Govt. of India, through the National Post Doctoral Fellowship (NPDF) (File No. PDF/2023/000161).

\bibliography{references}
\end{document}